\documentclass[12pt]{article}
 \usepackage{epsfig}
 \def\be{\begin{equation}}
 \def\ee{\end{equation}}
 \def\bea{\begin{eqnarray}}
 \def\eea{\end{eqnarray}}
 \usepackage{graphicx}

 \catcode`\@=11
 \def\lsim{\mathrel{\mathpalette\@versim<}}
 \def\gsim{\mathrel{\mathpalette\@versim>}}
 \def\@versim#1#2{\vcenter{\offinterlineskip
 \ialign{$\m@th#1\hfil##\hfil$\crcr#2\crcr\sim\crcr } }}
 \catcode`\@=12

 \catcode`@=12
\begin{document}
 \thispagestyle{empty}
 \begin{flushright}
 UCRHEP-T562\\
 January 2016\
 \end{flushright}
 \vspace{0.6in}
 \begin{center}
 {\LARGE \bf Soft $A_4 \to Z_3$ Symmetry Breaking\\
 and Cobimaximal Neutrino Mixing\\}
 \vspace{1.2in}
 {\bf Ernest Ma\\}
 \vspace{0.2in}
 {\sl Physics \& Astronomy Department and Graduate Division,\\ 
 University of California, Riverside, California 92521, USA\\}
 \vspace{0.1in}
 {\sl HKUST Jockey Club Institute for Advanced Study,\\ 
 Hong Kong University of Science and Technology, Hong Kong, China\\}

 \end{center}
 \vspace{1.2in}

 \begin{abstract}\
 I propose a model of radiative charged-lepton and neutrino masses with 
 $A_4$ symmetry.  The soft breaking of $A_4$ to $Z_3$ lepton triality is 
 accomplished by dimension-three terms.  The breaking of $Z_3$ 
 by dimension-two terms allows cobimaximal neutrino mixing ($\theta_{13} 
 \neq 0$, $\theta_{23} = \pi/4$, $\delta_{CP} = \pm \pi/2$) to be realized 
 with only very small finite calculable deviations from the residual $Z_3$ 
 lepton triality.  This construction solves a long-standing technical problem 
 inherent in renormalizable $A_4$ models since their inception.
 \end{abstract}

 \newpage
 \baselineskip 24pt

 For the past several years, some new things have been learned regarding 
 the theory of neutrino flavor mixing. (1) Whereas the choice of symmetry, 
 for example $A_4$~\cite{mr01,m02,bmv03}, 
 and its representations are obviously important, the breaking of this 
 symmetry into specific residual symmetries, for example 
 $A_4 \to Z_3$ lepton triality~\cite{m10,cdmw11}, is actually more important. 
 (2) A mixing pattern may be obtained~\cite{m04} independent of the masses 
 of the charged leptons and neutrinos. (3) The clashing of residual symmetries 
 between the charged-lepton, for example $A_4 \to Z_3$, 
 and neutrino, for example $A_4 \to Z_2$, sectors is technically very 
 difficult to maintain~\cite{af05}. (4) The essential incorporation of $CP$ 
 transformations~\cite{gl04,mn12} may be the new 
 approach~\cite{m15,cyd15,jp15,hrx15,h15,m16} which will lead to an improved 
 understanding of neutrino flavor mixing.

 In this paper, a model of radiative charged-lepton and neutrino masses 
 is proposed with the following properties. (1) The masses are generated 
 in one loop through dark matter~\cite{m06}, i.e. particles distinguished 
 from ordinary matter by an exactly conserved dark symmetry.  This is the 
 so-called scotogenic mechanism. 
 (2) The symmetry $A_4 \times Z_2$ 
 is imposed on all dimension-four terms of the renormalizable Lagrangian 
 with particle content given in Table 1. (3) Dimension-three terms break 
 $A_4 \times Z_2$, but all such terms respect the residual $Z_3$ lepton 
 triality. 
 (4) Dimension-two terms break $Z_3$, which is nevertheless retained 
 in dimension-three (and dimension-four) terms with only finite calculable 
 deviations. This solves the problem of clashing residual symmetries. 
 (5) The proposed specific model results in cobimaximal~\cite{m16} neutrino 
 mixing ($\theta_{13} \neq 0$, $\theta_{23} = \pi/4$, $\delta_{CP} = \pm 
 \pi/2$), which is consistent with the present data~\cite{pdg2014,t2k15}.  
 It is also 
 theoretically sound, because the residual $Z_3$ is protected, unlike 
 previous proposals.  Cobimaximal mixing becomes thus a genuine prediction, 
 robustly supported in the context of a complete renormalizable theory of 
 neutrino mass and mixing.

 \begin{table}[htb]
 \begin{center}
 \begin{tabular}{|c|c|c|c|c|}
 \hline
 particles & dark $U(1)_D$ & dark $Z_2$ & flavor $A_4$ & $Z_2$ \\ 
 \hline
 $(\nu,l)_L$ & 0 & + & 3 & + \\ 
 $l_R$ & 0 & + & 3 & $-$ \\ 
 $(\phi^+,\phi^0)$ & 0 & + & 1 & + \\ 
 \hline
 $N_{L,R}$ & 1 & $+$ & 3 & + \\ 
 $(\eta^+,\eta^0)$ & 1 & $+$ & 1 & + \\ 
 $\chi^+$ & 1 & $+$ & 1 & $-$ \\ 
 \hline
 $(E^0,E^-)_{L,R}$ & 0 & $-$ & 1 & + \\ 
 $F^0_{L}$ & 0 & $-$ & 1 & + \\ 
 $s$ & 0 & $-$ & 3 & + \\ 
 \hline
 \end{tabular}
 \caption{Particle content under $U(1)_D \times Z_2 \times A_4 \times Z_2$.}
 \end{center}
 \end{table}

 The dark $U(1)_D$ and $Z_2$ symmetries are assumed to be unbroken.  
 The other $Z_2$ symmetry 
 is used to forbid the dimension-four Yukawa couplings $\bar{l}_L l_R \phi^0$ 
 so that charged leptons only acquire masses in one loop as shown in Fig.~1.  
\begin{figure}[htb]
\vspace*{-3cm}
\hspace*{-3cm}
\includegraphics[scale=1.0]{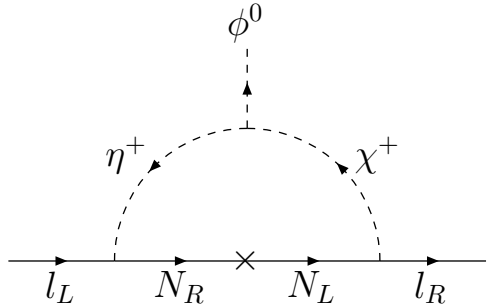}
\vspace*{-21.5cm}
\caption{One-loop generation of charged-lepton mass with $U(1)_D$ symmetry.}
\end{figure}
Whereas 
this $Z_2$ is respected by the dimension-four $\bar{l}_R N_L \chi^-$ terms, 
it is broken softly by the dimension-three trilinear $\eta^+ \chi^- \phi^0$ 
term to complete the loop.  This guarantees the one-loop charged-lepton mass 
to be finite.  Note that a dark $U(1)_D$ symmetry~\cite{m14,mpr13} is 
supported here with $\chi^+$, $(\eta^+,\eta^0)$, and $N_{L,R}$ all 
transforming as 1 under $U(1)_D$.  
The dimension-three soft terms $\bar{N}_L N_R$ are assumed to break $A_4$ 
to $Z_3$ through the well-known unitary matrix~\cite{mr01,c78,w78} 
$U_\omega$, i.e.
\begin{equation}
{\cal M}_N = U_\omega^\dagger \pmatrix{m_{N_1} & 0 & 0 \cr 0 & m_{N_2} & 0 \cr 
0 & 0 & m_{N_3}} U_\omega,
\end{equation}
where 
\begin{equation}
U_\omega = {1 \over \sqrt{3}} \pmatrix{1 & 1 & 1 \cr 1 & \omega & \omega^2 
\cr 1 & \omega^2 & \omega}.
\end{equation}
In the $A_4$ limit, ${\cal M}_N$ is proportional to the identity matrix. 
With three different mass eigenvalues, the residual symmetry is $Z_3$ 
lepton triality.  Let the $(\eta^+,\chi^+)$ mass eigenvalues be $m_{1,2}$ 
with mixing angle $\theta$, then each lepton mass is given by~\cite{m14}
\begin{equation}
m_l = {f_L f_R \sin \theta \cos \theta m_N \over 16 \pi^2} 
[F(x_1) - F(x_2)],
\end{equation}
where $F(x) = x \ln x/(x-1)$, with $x_{1,2} = m^2_{1,2}/m^2_N$.

The dark $U(1)_D$ symmetry forbids the quartic scalar term 
$(\Phi^\dagger \eta)^2$, so that a neutrino mass is not generated as 
in Ref.~\cite{m06}.  It comes instead from Fig.~2, where the scalars 
$s_{1,2,3}$ are assumed real~\cite{m15,fmp14,mnp15} to enable cobimaximal 
mixing, hence a separate dark $Z_2$ symmetry 
is required. 
\begin{figure}[htb]
\vspace*{-3cm}
\hspace*{-3cm}
\includegraphics[scale=1.0]{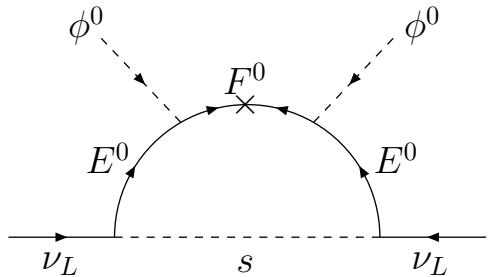}
\vspace*{-21.5cm}
\caption{One-loop generation of neutrino mass from $s$.}
\end{figure}
Let the $\bar{F}_L E_R$ mass term be $m_D$ and assumed to be 
much smaller than $m_E,m_F$, then each neutrino mass is given by
\begin{equation}
m_\nu = {h^2 m_D^2 m_F \over 16 \pi^2 (m_F^2 - m_s^2)}[G(x_F) - G(x_s)],
\end{equation}
where
\begin{equation}
G(x) = {x \over 1-x} + {x^2 \ln x \over (1-x)^2},
\end{equation}
with $x_F = m_F^2/m_E^2$, $x_s = m_s^2/m_E^2$. 
The dimension-two $s_i s_j$ terms are allowed to break $Z_3$ arbitrarily. 
However, since this mass-squared matrix is real, it is diagonalized by an 
orthogonal matrix ${\cal O}$, hence the neutrino mixing matrix is 
given by~\cite{m15,fmty00,mty01}
\begin{equation}
U_{l \nu} = U_\omega {\cal O},
\end{equation}
resulting in $U_{\mu i} = U^*_{\tau i}$, thus guaranteeing cobimaximal 
mixing: $\theta_{13} \neq 0$, $\theta_{23} = \pi/4$, $\delta_{CP} = \pm 
\pi/2$.  

In a previous proposal~\cite{m15}, instead of Fig.~1, the radiative 
charged-lepton masses also come from scalars, i.e. $x_i^+ \sim \underline{3}$, 
$y_i^+ \sim \underline{1}, \underline{1}', \underline{1}''$ under $A_4$. 
The $A_4 \to Z_3$ breaking is accomplished by rotating $x_i^+$ through 
$U_\omega$ so that $x_{1,2,3}^+$ now correspond to $y_{1,2,3}^+$ under 
$Z_3$, and allowing the $(x_1,y_1)$, $(x_2,y_2)$, $(x_3,y_3)$ sectors to 
have separate arbitrary masses.  Now the quartic scalar coupling 
$(x_1^+ s_1 + x_2^+ s_2 + x_3^+ s_3)(x_1^- s_1 + x_2^- s_2 + x_3^- s_3)$ 
is allowed under $A_4$.  If the $s_i s_j$ mass-squared terms break $Z_3$ 
as in Fig.~2, then the $s_1 s_2 (x_1^+ x_2^- + x_2^+ x_1^-)$ term from the 
above will induce a quadratic $x_1 x_2$ term as shown in Fig.~3.  
\begin{figure}[htb]
\vspace*{-4cm}
\hspace*{-3cm}
\includegraphics[scale=1.0]{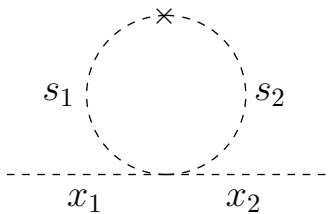}
\vspace*{-21.5cm}
\caption{One-loop generation of $x_1 x_2$ term from $s_1 s_2$ term.}
\end{figure}
Whereas 
this diagram is not quadratically divergent, it is still logarithmically 
divergent.  This means a counterterm is required for 
$x_1^+ x_2^- + x_2^+ x_1^-$, thereby invalidating the $Z_3$ residual 
symmetry necessary to derive $U_\omega$ and thus Eq.~(6).

In this proposal, the $A_4 \to Z_3$ breaking comes from $\bar{N}_L N_R$, 
with the Dirac fermions $N_{1,2,3}$ distinguished from one another by the 
residual $Z_3$ lepton triality through $U_\omega$ as shown in Eq.~(1).  
The soft breaking of $Z_3$ by $s_1 s_2$ induces only a finite two-loop 
correction to the $N_1 - N_2$ wavefunction mixing as shown in Fig.~4.
\begin{figure}[htb]
\vspace*{-4cm}
\hspace*{-3cm}
\includegraphics[scale=1.0]{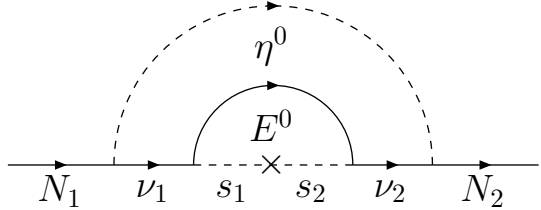}
\vspace*{-21.5cm}
\caption{Two-loop $N_1-N_2$ mixing from $s_1 s_2$ breaking of $Z_3$.}
\end{figure}
Therefore this construction solves a long-standing technical problem in 
renormalizable theories of $A_4$ flavor mixing.  To summarize, (1) $A_4$ 
is respected by all dimension-four terms; (2) $Z_3$ is respected by all 
dimension-three terms; (3) $Z_3$ is broken arbitrarily by dimension-two 
terms to allow cobimaximal mixing according to Eq.~(6); (4) the $s_i s_j$ 
terms generate very small finite radiative corrections to $Z_3$ breaking 
in the dimension-three terms, justifying the use of $U_\omega$ 
to obtain Eq.~(6).

As for dark matter, there are in principle two stable components:  the 
lightest $N$ with $U(1)_D$ symmetry and the lightest $s$ with $Z_2$ 
symmetry.  Whereas $N$ has only the allowed $\bar{N}_R (\nu_L \eta^0 - l_L 
\eta^+)$ interactions, $s$ has others, i.e. $s^2 \Phi^\dagger \Phi$, 
$s^2 \eta^\dagger \eta$, $s^2 \chi^+ \chi^-$, as well as 
$s(\bar{\nu}_L E_R^0 + \bar{l}_L E^-_R)$.  Their interplay to make up 
the total correct dark-matter relic abundance of the Universe and how 
they may be detected in underground direct-search experiments require 
further study. 

An immediate consequence of radiative charged-lepton mass is that the 
Higgs Yukawa coupling  $h \bar{l} l$ is no longer exactly 
$m_l/(246~{\rm GeV})$ as predicted by the standard model, as studied in 
detail already~\cite{fm14,fmz15}.  Because of the $Z_3$ lepton triality, 
large anomalous muon magnetic moment may be accommodated while 
$\mu \to e \gamma$ is suppressed~\cite{fmz15}.

In conclusion, cobimaximal neutrino mixing $(\theta_{13} \neq 0, \theta_{23} = 
\pi/4, \delta_{CP} = \pm \pi/2)$ is achieved rigorously in a renormalizable 
model of radiative charged-lepton and neutrino masses.  The key is the soft 
breaking of $A_4$ to $Z_3$ by dimension-three terms, so that the subsequent 
breaking of $Z_3$ by dimension-two terms only introduces very small finite 
corrections to the $U_\omega$ transformation needed to obtain cobimaximal 
mixing as given by Eq.~(6).

\medskip
This work is supported in part 
by the U.~S.~Department of Energy under Grant No.~DE-SC0008541.

\baselineskip 16pt
\bibliographystyle{unsrt}

\end{document}